\documentstyle[aps,epsfig,prl,multicol]{revtex}

\begin{document}
\title{
Finite thermal conductivity in 1D models having zero Lyapunov exponents}
\author{
Baowen Li$^1$, Lei Wang$^{2}$, and
Bambi Hu,$^{2,3}$
}
\address{
$^{1}$  Department of Physics, National University of Singapore,
117542 Singapore \\
$^{2}$Department of Physics and Center for Nonlinear Studies, Hong
Kong Baptist University, China\\
$^{3}$ Department of Physics and Texas Center for Superconductivity, University of Houston, Houston TX
77204-5506
}
\date{Phys. Rev. Lett {\bf 88}, 223901 (2002)}
\maketitle

\begin{abstract}
Heat conduction in  three types of 1D channels are studied.  The
channels consist of two parallel walls, right triangles as
scattering obstacles, and noninteracting particles. The triangles
are placed along the walls in three different ways: (a) periodic,
(b) disordered in height, and (c) disordered in position. The
Lyapunov exponents in all three models are zero because of the
flatness of triangle sides. It is found numerically that the
temperature gradient can be formed in all three channels, but the
Fourier heat law is observed only in two disordered ones. The
results show that there might be no direct connection between
chaos (in the sense of positive Lyapunov exponent) and the normal
thermal conduction.
\end{abstract}

\begin{multicols}{2}

Recent years has witnessed an increasing attention
to the establishment of a connection between macroscopic phenomena such as transport coefficient
and microscopic chaos\cite{chaos98,dorfman99,cohen}. A direct mathematical derivation has been proved to be very difficult, and only very simple model can such an approach be established\cite{LS78-82}, we have to rely on massive numerical simulations.
There have been a large number of numerical works on heat conduction in 1D systems\cite{Casati84,FPU,HLZ98,FHLZ98,Zhang00,Bishop00,Hatano99,Dhar99,Casati99,Prosen00,Period00,LZH01,Aoki01,MLL01,Garrido01}
aim to understand what are the necessary and sufficient conditions for a Hamiltonian system to obey the Fourier heat conduction law. It is found that an on-site potential is sufficient for  a 1D lattice model to have a finite thermal conductivity\cite{HLZ98}.

Albeit many progress achieved, open questions remain (see recent review\cite{Lebowitz00}).
For example, in connecting the normal heat conduction with the underlying dynamics, some contradictions exist. On the one hand, some models like the ding-a-ling model\cite{Casati84} and the Lorentz gas model (with periodic and/or disordered disks) showing exponential instability, thus a positive Lyapunov exponent, have a normal heat conduction\cite{LS78-82,Casati84,Casati99}. On the other hand, the Fermi-Pasta-Ulam (FPU) model has a divergent thermal conductivity\cite{FPU} even though it has positive Lyapunov exponents. Therefore, what's a role does chaos (in the sense of positive Lyapunov exponent) play in the normal heat conduction is still a unsolved problem and deserves further investigation.

In this Letter, we study this problem in a series of 1D models having zero Lyapunov exponents. Our models are variants of the Ehrenfest model\cite{Ehrenfest} and thus called ``Ehrenfest gas channels". The channel consists of two parallel walls, a series of isosceles right triangles with hypotenuse along the parallel walls, and noninteracting particles. The two ends of the channel are put in contact with heat baths.  By placing the triangles in different ways, we obtain different types of channels.

The Ehrenfest model differs from the Lorentz gas model in underlying dynamics. The collisions of the particles with the circles in the Lorentz gas lead to exponential separation of nearby trajectories, thus a positive Lyanpunov exponent, whereas collisions with the squares in the Ehrenfest model lead to linear separation of nearby trajectories, thus a zero Lyapunov exponent.

{\it Channel with periodic structure} In this channel, the right triangles are placed periodically, namely, in each cell, we have two triangles, one on the bottom wall, the other on the top. The triangles are placed at the position of $x=1,3,\cdots$ (arbitrary unit). The model geometry is shown in Fig. 1(a). The channel of length $N$ is $N$'th repetition of the cell. Two heat baths with temperature $T_+$ and $T_-$ are attached to the left and right ends of the channel, respectively . The heat bath has simple velocity distribution $P_T(v) = \delta(v-\sqrt{2T})$. It can be proved that the form of heat baths does not affect the transport behavior in our systems.

To compute temperature field at a stationary state, we calculate time averages by dividing the configuration space into a set of boxes ${C_i}$ \cite{Casati99}. The time spent within a box in the $j$th visit is denoted by $t_j$ and the total  number of crossings of a box $C_j$ during the simulation is $M$. The  temperature field is defined by\cite{Casati99}
\begin{equation}
T_{C_i} = \langle E \rangle_{C_i} = \frac{\sum_j^M t_jE_j(C_i)}{\sum_j^M t_j}.
\end{equation}
Then it is projected on $x$ direction (the transport direction).
The heat flux is calculated by the change of energy carried through to the left and right ends by the particles,
\begin{equation}
J = \frac{1}{t_M}\sum_{j=1}^{M} \Delta E_ j,
\end{equation}
where $\Delta E_j= (E_{in} - E_{out})_j$ is the energy change at the $j$th collision with a heat bath, $t_M$ is the total time spent for $M$ such collisions.

In numerical simulation, we compute the flux for a single particle $J_1$. The scaled heat flux is $J_N(N) = NJ_1(N)$\cite{Casati99}, where $N$ is the number of the cells. Each cell has length $a$, thus the channel has length $L=Na$.

\begin{figure}
\epsfxsize=8.cm \epsfbox{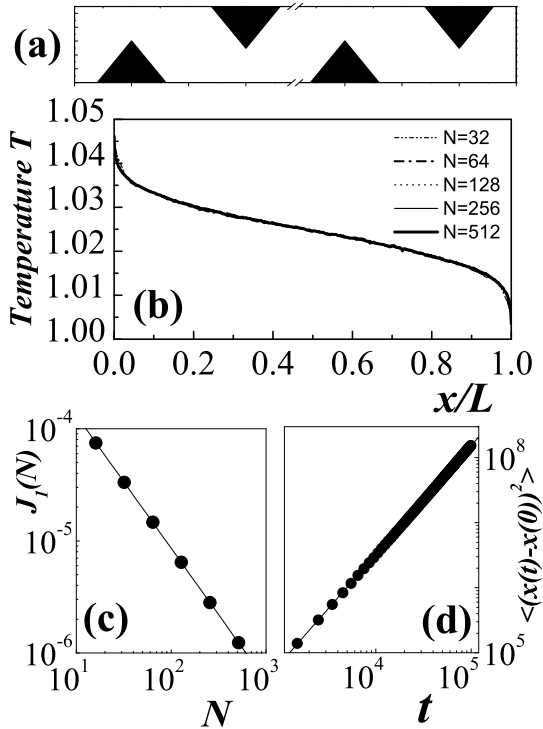} \vspace{0cm} \narrowtext
\caption{{\it Periodic channel}. (a) Geometry. (b) Temperature
profile. (c) Heat flux $J_1(N)$ versus $N$. (d) $\langle (x(t) -
x(0))^2 \rangle$ versus $t$. In (c) and (d) the bullets are the
numerical data and the solid lines are the best fit ones. The
width of the channel is 1.1 (arbitrary unit) and the height of
right triangle is 0.6. } \label{Model1}
\end{figure}

In spite of the jumps at both ends the temperature gradient is well established and
scales as $dT/dx\propto N^{-1}$ as shown in Fig. 1(b). The heat flux $J_1(N)$ is found to be
\begin{equation}
J_1(N) = AN^{\alpha},
\label{J1N}
\end{equation}
with $\alpha =-1.186\pm0.002$. The thermal conductivity $\kappa = -\frac{J_N(N)}{dT/dx} \propto N J_N \sim N^{0.81}$, which is divergent as one goes to the thermodynamic limit ($N\to \infty$).

To understand this divergent behavior, we study the transport property of the particles in the channel quantified by the mean square displacement $\langle (x(t) - x(0))^2 \rangle$. An ensemble of particles ($10^5$) with the same amplitude of velocity $(=1)$ are injected into the channel in random directions. The best fit for the asymptotic behavior gives rise to
\begin{equation}
\langle (x(t) - x(0))^2 \rangle = D t^{\beta},
\label{Diffusion}
\end{equation}
with $\beta=1.672\pm 0.003$ (Fig 1(d)). This means that the transport along $x$ direction is neither a ballistic one ($\beta=2$) nor a diffusive one ($\beta=1$). This super diffusion is responsible for the divergent thermal conductivity. It may also be the reason for the jumps near the channel ends in the temperature profiles.  Such jumps have been
observed in the FPU model\cite{FPU,HLZ98} and attributed to the soliton alike excitations\cite{FHLZ98,Zhang00}. A quantitative analysis has been done by Aoki and Kusnezov\cite{Aoki01} more recently.

When our model is compared with the Lorentz gas channel\cite{Casati99}, it is intuitive to attribute the divergent thermal conductivity to the zero Lyapunov exponent. To clarify this point, we modify the channel slightly in two ways: (a) make the height of triangles random; (b) put the triangles in random position along the transport direction. The Lyapunov exponent in both variants remain zero because of the flatness of the triangle sides.

{\it Channel with right triangles of random heights}
 The height of the triangle is given by
\begin{equation}
h_i = h_0 + d*R_i,\quad i=1,2,\cdots,2N
\end{equation}
where $\{R_i\}$ are random numbers uniformly distributed in the interval $[-1,1]$, $d$ is the magnitude of disorder. $h_i<H$, where $H$ is the width of the channel.
Figure 2(a) shows the geometry of the channel. In this Letter, we take $H=1.1$, $h_0=0.6$, and $d \in [0,0.4]$.

In our calculations, the temperature and heat flux are averaged over 100 disorder realizations and compared with that one from averaged over 1000 realizations, the difference is found to be indistinguishable.

Figure 2(b) shows the temperature profile for $d =0.4$. It is a straight line with gradient $dT/dx = -0.05/N$. The heat flux $J_1(N)$ is shown in Fig. 2(c). The best fit gives rise to a slope $\alpha = -1.992 \pm 0.018$. Therefore $J_N(N)\sim N^{-1}$. The thermal conductivity $\kappa = - J_N(N)/(dT/dx)$ is an $N$ independent constant, the Fourier
law is thus justified. To see how the heat conduction changes with disorder, we calculate the exponent $\alpha$ for different values of $d$ by fixing the channel length. The results are shown in Fig.2(d). The bullets represent the $\alpha$ values obtained from the best fit with Eq.(\ref{J1N}) by using
$N \in [16, 512]$. It shows that, for a disordered channel of finite length, the heat conduction obeys Fourier law when the disorder amplitude is large enough. In principle, in the thermodynamic limit, any infinitesimal disorder will cause a diffusive transport, thus a normal thermal conduction. This is demonstrated by the case with $d=0.0125$, $\alpha=-1.724\pm 0.012$
from the data $N \in [16, 512]$ which is far from the normal thermal conduction, however,  $\alpha=-1.999\pm 0.011$ from $N\in [1024, 32768]$ (the star in Fig. 2(d).) showing  a normal thermal conductivity. This is similar to the mass disordered lattice model \cite{LZH01}.

\begin{figure}
\epsfxsize=8.cm
\epsfbox{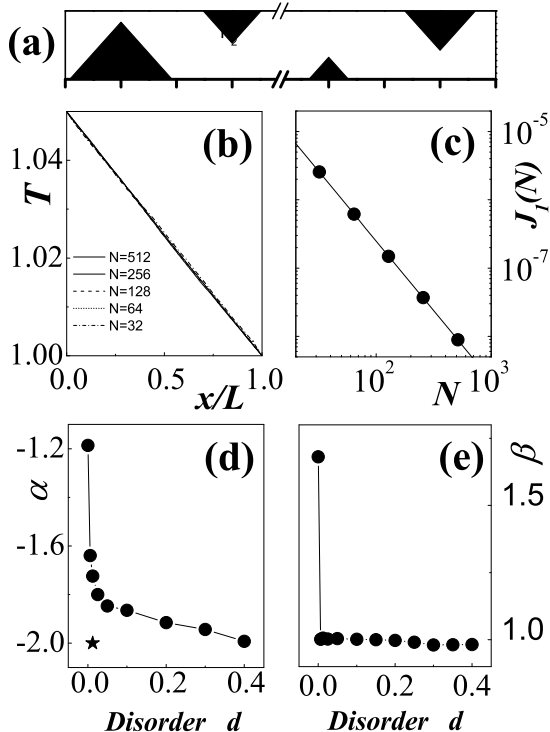}
\vspace{0cm}
\narrowtext
\caption{{\it Height disordered channel}. (a) Channel geometry. (b) Temperature profile. (c) Heat flux $J_1(N)$ versus $N$.  (d) $\alpha$ versus $d$. (e) $\beta$ versus $d$. $d=0.4$ in (b) and (c). The solid lines in (d) and (e) are drawn to guide eyes.
}
\label{Model2}
\end{figure}

We compute $\langle(x(t) -x(0))^2\rangle $ and find that for all values of disorder, it can be best fitted by $ Dt^{\beta} $ asymptotically.  $\beta$ as a function of the disorder $d$ is plotted in Fig. 2(e). It is seen that for any finite value of $d$, the slope $\beta$ is  very close to unity, which means that the particles moves diffusively in the channel, thus the heat conduction in this channel obeys Fourier law.

Thermal conductivity $\kappa$ versus temperature $T_0= (T_+ + T_-)/2$ is plotted in Fig. 3(a). It is found that $\kappa \sim T_0^{\gamma}$, and the best fit gives rise to $\gamma = 0.501\pm 0.002$. The normalized temperature profile $T^*(x)=T(x)/T_0$ is shown in Fig. 3(b), which indicate that $dT/dx = -0.02T_0/L$.

{\it Channel with triangles at random positions} The position of the triangle is made random, namely, $x_i=d*R_i$, where $x_i$ is the position away from the periodic structure shown in Fig. 1(a).  Figure 4(a) is the schematic illustration of the geometry.
The numerical simulations show that the temperature gradient is well established and is similar to Fig. 2(b). The heat flux $J_1(N)$ is described by Eq. (\ref{J1N}). We plot the exponent $\alpha$ versus $d$ in Fig. 4(b). It tells us the trend to normal thermal conductivity ($\alpha =-2$) in a long length limit (star) and a large disorder limit (bullets).

As an independent check, we also calculate the integral of the current-current correlation function in the Green-Kubo formula. The integral is found to be convergent in cases with disorder but divergent in the case with periodic geometry shown in Fig. 1(a).

\begin{figure}
\epsfxsize=8.cm
\epsfbox{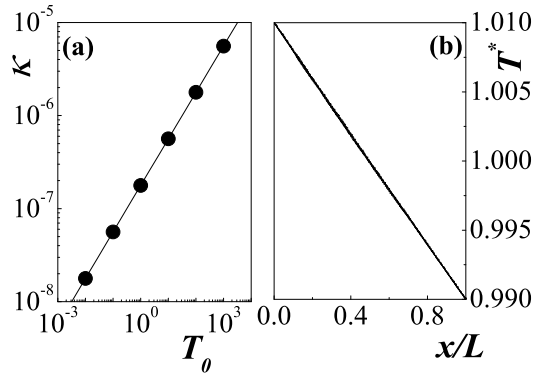}
\vspace{0cm}
\narrowtext
\caption{{\it Height disordered channel}. (a) Thermal conductivity $\kappa$ versus temperature $T_0=(T_+ + T_-)/2$.
(b) The normalized temperature profile ($T^* = T/T_0$) for six different temperature scales. $T_0= 0.01, 0.1, 1, 10, 100$, and $1000$, respectively. Disorder $d=0.4$.
}
\label{KappaT}
\end{figure}

{\it Theoretical analysis} Suppose the path length distribution of particles from left to right (or vice versa) is $f_L(l)$ in a channel of length $L$, namely, there are $\delta n$ particles whose path length lies in the interval $[l,l+dl]$. $\delta n/n = f_L(l)dl$. $f_L(l)$ is determined merely by the structure and the length of the channel.  Two heat baths of temperature $T_L$ and $T_R$ are put to the left and right ends, respectively. In a time period of $t$ there are $n$ particles exchanged between two heat baths. The total time spent to reach the right heat bath from the left one is
$
t_{LR} = n\langle l\rangle\int_0^{\infty}\frac{1}{v}P_{T_1}(v)dv,
$
where $\langle l\rangle=\int_0^{\infty}l f_L(l)dl$ is the average path length from the left heat bath to the right one, and
$P_T(v)$ the velocity distribution function of heat bath at temperature $T$.
Similarly, the total time of $n$ particles from the right bath to the left one is
$t_{RL}= n \langle l\rangle \int_0^{\infty}\frac{1}{v}P_{T_2}(v)dv$.
The total time is $t= t_{LR}+t_{RL}$.  The energy exchange between two heat baths is $
E = n\int_0^{\infty}\frac{v^2}{2}(P_{T_1}(v) - P_{T_2}(v))dv = n(T_1-T_2)$,
the heat flux for the channel of length $L$ per particle is thus given by:
\begin{equation}
J_1(L) =\frac{E}{t} = \frac{T_1-T_2}{\langle l\rangle \int_0^{\infty}\frac{1}{v}(P_{T_1}(v) + P_{T_2}(v))dv}.
\label{TheoJ1}
\end{equation}

From Eq(\ref{TheoJ1}), it can be seen that whether the heat conduction obeys the Fourier law or not does not depend on the types of heat bath, it depends only on $\langle l\rangle$ - the transport property. For instance, if the system is diffusive, then  $\langle l\rangle \propto L^2$ and the heat flux $J_1(L) \propto L^{-2}$. This is what we see in numerical calculations (Fig. 2(c)). Moreover, for a given geometry, i.e. $\langle l\rangle$ is  determined, the heat flux and heat conductivity are determined by the property of the heat baths. If we change the temperature of  heat baths $q$ times, then for the simple heat bath we used and the Gaussian heat bath, it can be shown that, the heat flux changes $q^{3/2}$ times.  Because the temperature gradient $dT/dx= \mbox{Const.} T_0$ (see Fig. 3(b)), thus the thermal conductivity $\kappa$ changes with temperature $T_0^{1/2}$, this agrees with our numerical finding in Fig. 3(a).

\begin{figure}
\epsfxsize=8.cm
\epsfbox{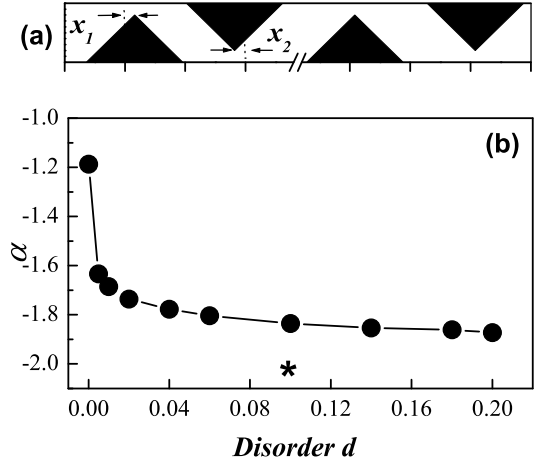}
\vspace{0cm}
\narrowtext
\caption{{\it Position disordered channel}. (a) Geometry.
(b) $\alpha$ versus $d$. $\{x_i\}=d*R_i$, $d$ is disorder. The solid line in (b) is drawn to guide eyes. In (b) the bullets are the data from best fit by using $N\in [32,1024]$, while ``*" at $d =0.1$ is the one by using $N\in[8192,32768]$.}
\label{Model3}
\end{figure}

In summary, we have studied heat conduction in three different 1D Ehrenfest channels.  The temperature gradient can be formed in all cases. However, a finite thermal conductivity can be reached only when the disorder (either in position or in height) exists. As the Lyapunov exponents are zero in our model, we thus conclude that the finite thermal conductivity might have nothing to do with the underlying dynamics. Most recent study on heat conduction in channels with irrational triangles supports this argument\cite{Li02}.

\bigskip
BL would like to thank G Casati for useful discussions, he
was supported in part by Academic Research Fund of NUS.  LW and BH were supported in part by Hong
Kong RGC, the HKBU's FRG, and the Texas Center for Superconductivity.

$^*$ Author to whom correspondence should be addressed. Email:
phylibw@nus.edu.sg

\end{multicols}

\begin{thebibliography}{9}

\bibitem{chaos98}
``{\it Focus issue: Chaos and Irreversibility}'', Chaos {\bf 8},
(1998); J. Lebowitz, Physica A {\bf 263}, 516
(1999); I. Prigogine, {\it ibid.}  p. 528; D. Ruelle, {\it
ibid.}  p. 540;
P Gaspard {\it et al}, Nature {\bf 394}, 865 (1998); C. P. Dettermann, E. G. D. Cohen, and H. van Beijeren, {\it ibid.} {\bf 401}, 875 (1999).



\bibitem{dorfman99}J R Dorfman,
"{\it An Introduction to Chaos in Nonequilibrium Statistical Mechanics}". (Cambridge University Press, 1999).

\bibitem{cohen}
C. P. Dettermann and E. G. D. Cohen, J. Stat. Phys. {\bf 101}, 775 (2000); {\bf 103}, 589 (2001).


\bibitem{LS78-82}
J. L. Lebowitz and H. Spohn, J. Stat. Phys. {\bf 19}, 633 (1978); {\bf 28}, 539 (1982).


\bibitem{Casati84}
G. Casati, J. Ford, F. Vivaldi, and W. M. Visscher,
Phys. Rev. Lett. {\bf 52}, 1861 (1984);
T. Prosen and M. Robnik, J. Phys. A {\bf 25}, 3449
(1992); D. J. R. Mimnagh and L.E. Ballentine, Phys. Rev. E {\bf 56}, 5332
(1997); H. A. Posch and Wm. G. Hoover, {\it ibids.} {\bf 58} 4344 (1998).

\bibitem{FPU}
H. Kaburaki and M. Machida, Phys. Lett. A {\bf 181}, 85
(1993); S. Lepri, R. Livi, and A Politi, Phys. Rev. Lett. {\bf 78},
1896 (1997)

\bibitem{HLZ98}
B. Hu, B. Li, and H. Zhao, Phys. Rev. E {\bf 57}, 2992 (1998); B. Hu, B. Li, and H. Zhao, Phys. Rev. E {\bf 61}, 3828(2000);

\bibitem{FHLZ98}
A. Fillipov, B. Hu, B. Li, and A. Zeltser, J. Phys. A {\bf
31}, 7719 (1998).

\bibitem{Zhang00}
F. Zhang, D. J. Isbister, and D. J. Evans, Phys. Rev. E {\bf 61}, 3541 (2000).

\bibitem{Bishop00}
G. P. Tsironis, A. R. Bishop, A. V. Savin, and A. V. Zolotaryuk, Phys. Rev. E. {\bf 60}, 6610 (2000); A Aoki and D.Kusnezov, Phys. Lett. A {\bf  265}, 250 (2000).

\bibitem{Hatano99}
T. Hatano, Phys. Rev. E {\bf 59}, R1 (1999).

\bibitem{Dhar99}
A. Dhar and D. Dhar, Phys. Rev. Lett. {\bf 82}, 480 (1999); A. Dhar, Phys. Rev. Lett. {\bf 86}, 3554 (2001).

\bibitem{Casati99}
D.  Alonso, R. Artuso, G. Casati, and I. Guarneri, Phys. Rev. Lett. {\bf
82}, 1859 (1999).

\bibitem{Prosen00}
T. Prosen and D. K. Campbell, Phys. Rev. Lett. {\bf 84}, 2857 (2000).

\bibitem{Period00}
C. Giardin'a, R. Livi, A. Politi, and M. Vassalli,
Phys. Rev. Lett. {\bf 84}, 2144 (2000); O.V. Gendelman and A. V. Savin, {\it ibid.} {\bf 84}, 2381 (2000).

\bibitem{LZH01}
B. Li, H. Zhao, and B. Hu, Phys. Rev. Lett. {\bf 86}, 63 (2001);
B. Li, H. Zhao, and B. Hu, Phys. Rev. Lett. {\bf 87}, 069402 (2001).

\bibitem{Aoki01}
K. Aoki, and D. Kusnezov, Phys. Rev. Lett. {\bf 86}, 4029 (2001).

\bibitem{MLL01}
C. Mejia-Monasterio, H. Larralde, and F. Leyvraz, Phys. Rev. Lett. {\bf 86}, 5417 (2001).

\bibitem{Garrido01}
P. L. Garrido, P. I. Hurtado, and B. Nadrowski, Phys. Rev. Lett. {\bf 86}, 5486 (2001).

\bibitem{Lebowitz00} F Bonetto, J L Lebowitz, and L. Rey-Bellet, e-preprint, math-ph/0002052.

\bibitem{Ehrenfest}
P. and T. Ehrenfest, ``{\it The Conceptual Foundations of The Statistical Approach in Mechanics}", Dover Pub. Inc., New York (1959).

\bibitem{Li02}
B. Li, G. Casati, and J. Wang ``Heat conduction in linear mixing
systems", NUS-Preprint 2002.


\end{thebibliography}
\end{document}